\documentclass[aps,pra,twocolumn,showpacs,floatfix,nofootinbib,groupedaddress,superscriptaddress]{revtex4}
\usepackage{amssymb,graphics,graphicx,times,bm,bbm,multirow,color}

\begin{document}

\title{Environment-Assisted Quantum Transport}
\author{Patrick Rebentrost}
\affiliation{Department of Chemistry and Chemical Biology, Harvard University, 12 Oxford
St., Cambridge, MA 02138}
\author{Masoud Mohseni}
\affiliation{Department of Chemistry and Chemical Biology, Harvard University, 12 Oxford
St., Cambridge, MA 02138}
\author{Ivan Kassal}
\affiliation{Department of Chemistry and Chemical Biology, Harvard University, 12 Oxford
St., Cambridge, MA 02138}
\author{Seth Lloyd}
\affiliation{Department of Mechanical Engineering, Massachusetts Institute of Technology,
77 Massachusetts Avenue, Cambridge MA 02139}
\author{Al\'an Aspuru-Guzik}
\affiliation{Department of Chemistry and Chemical Biology, Harvard
University, 12 Oxford St., Cambridge, MA 02138}
\email{aspuru@chemistry.harvard.edu} 

\begin{abstract}
Transport phenomena at the nanoscale are of interest due to the presence of
both quantum and classical behavior. In this work, we demonstrate that
quantum transport efficiency can be enhanced by a dynamical interplay of the
system Hamiltonian with pure dephasing induced by a fluctuating
environment. This is in contrast to fully coherent hopping that leads to
localization in disordered systems, and to highly incoherent transfer that
is eventually suppressed by the quantum Zeno effect. We study these
phenomena in the Fenna-Matthews-Olson protein complex as a prototype for
larger photosynthetic energy transfer systems. We also show that disordered
binary tree structures exhibit enhanced transport in the presence of dephasing.
\end{abstract}

\volumeyear{year}
\volumenumber{number}
\issuenumber{number}
\eid{identifier}
\date{\today}
\startpage{1} \maketitle

The dynamical behavior of a quantum system can be substantially
affected by the interaction with a fluctuating environment. Noise and
decoherence collapse the quantum wavefunction, and
one might be lead to expect an inhibitory effect
on, for example, quantum transport involving coherent hopping of a (quasi-)
particle between localized sites.
One of the most important classes of quantum transport is the energy transfer in
molecular systems \cite{MayBook}, for example in the chromophoric
light-harvesting complexes \cite{Engel07,Lee07}. The role of the environment in chromophoric
systems \cite{Grover71,Yang02,Gilmore08} and model geometries \cite{Gaab04}
has been widely studied. The
Haken-Strobl model is used to describe Markovian bath fluctuations
\cite{Haken73,Gaab04,Leegwater96}.
Quantum transport can also be affected by the well-known
quantum localization \cite{Anderson58,Anderson78}. Energy mismatches in disordered materials
lead to destructive interference of the wavefunction and
subsequently to localization of the quantum particle. Specifically, it has
been argued that quantum localization can seriously limit computational
power and/or quantum walk properties in binary tree structures \cite{Keating07},
where an exponential speed-up over a
classical random walk can in principle be observed  \cite{Childs02}.
Generally, the overall effect of
environment and static disorder is expected to be negative. However, as we
demonstrate here, in a large variety of transport systems and under proper
conditions, the interaction with the environment can result in increased quantum transport
efficiency.

\begin{figure}[tbp]
\begin{center}
\includegraphics{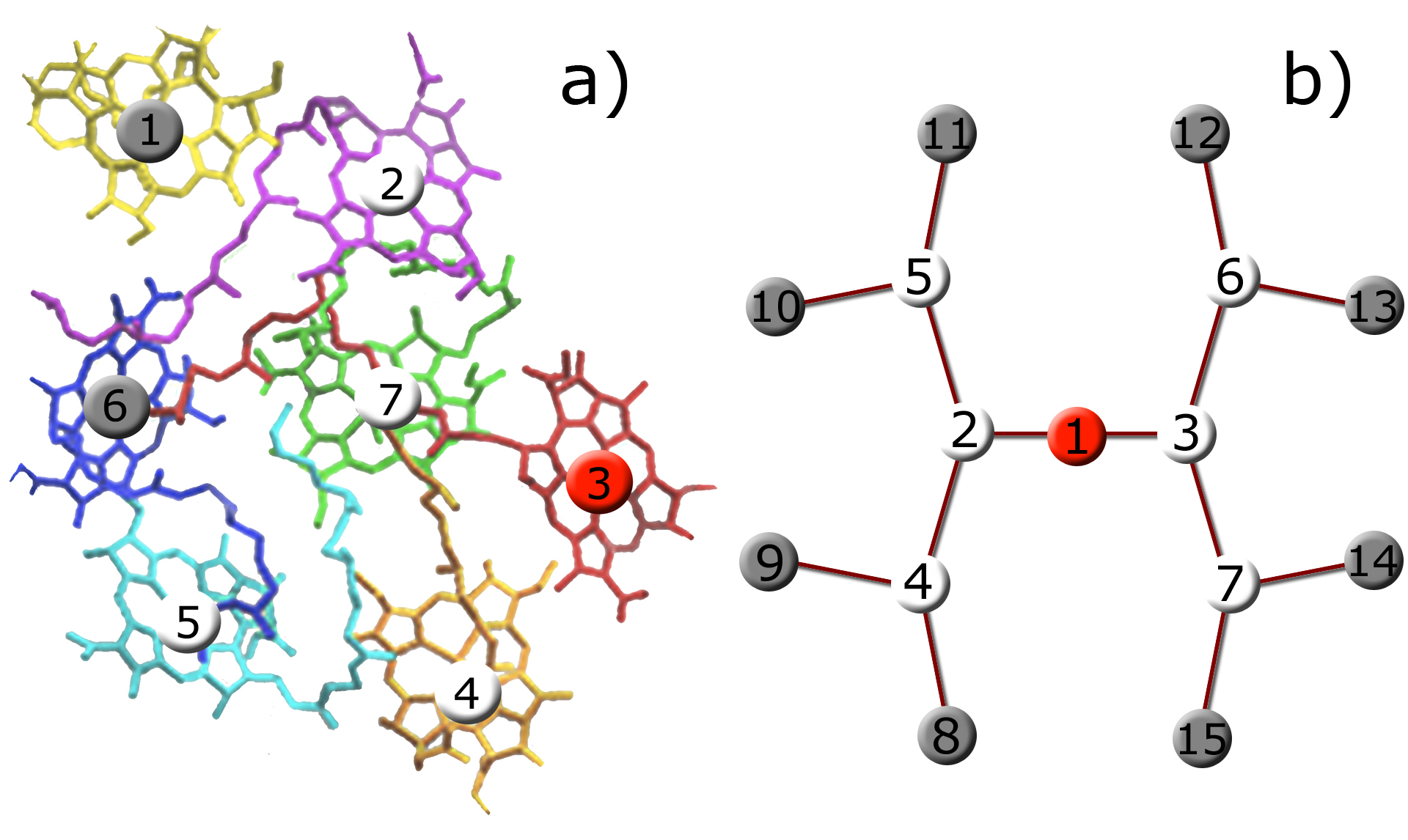}
\end{center}
\caption{Quantum transport occurs in natural and engineered systems,
for example: a) Energy transfer in photosynthetic complexes between
chlorophyll molecules, such as the Fenna-Matthews-Olson protein
complex, in which quantum coherence has been shown to play a
significant role in the exciton dynamics \cite{Engel07}. (b)
Transport of particles and/or information in artificial or
engineered systems described by a tight-binding Hamiltonian.
Here, a four generation binary tree is shown. In particular, an exponential speed-up in
reaching certain target sites (red) in these structures has been
proposed in the context of quantum walk algorithms \cite{Childs02}.
(The grey sites represent initial states for the quantum transport.)
} \label{figTransport}
\end{figure}

In chromophoric complexes,
an environment-assisted quantum walk approach within a Redfield model
involving relaxation and dephasing was
suggested to explain the high energy transfer efficiency \cite{Mohseni08}. This
approach was also used to quantify the percentage contributions of quantum
coherence and environment-induced relaxation to the overall efficiency \cite%
{Rebentrost08}. Phonon-enhanced transfer has recently been reported for two quantum
dots within the Redfield theory including relaxation \cite{Rozbicki08}.
Measurement of a single site
in a quantum-dot array leads to complete delocalization of electrons in
the one-dimensional Anderson model \cite{Gurvitz00}.
Noise-induced enhancement can also been seen in
stochastic resonance \cite{Gammaitoni98}, where the quantum system
is driven to a non-linear regime. The role of pure dephasing in
quantum localization was discussed in \cite{Logan87,Logan90,Parris88,Phillips93}.
Several authors investigated quantum entanglement in biological systems in the presence of
dissipative dynamics, reset mechanisms, and molecular oscillations \cite{Hartmann07,Cai08}.

Interactions of a quantum system with a thermal environment or measurement of an
observable can lead to the Zeno and anti-Zeno effects. In the Zeno
effect \cite{Misra77}, sometimes also termed the 'watchdog effect',
repeated, fast measurement suppresses the free evolution governed by
Schr\"odinger's equation. The quantum system remains in an
eigenstate of the measurement operator. The anti-Zeno effect
describes the opposite scenario \cite{Kofman00,Facchi01,Erez08}.
Here, such interactions, if well timed, accelerate certain
processes, such as the decay of an unstable state, compared to the
unperturbed case.

In this work, we study the interaction between \emph{pure dephasing
noise} and coherent dynamics that can result in greatly increased
efficiency of transport. Unlike stochastic resonance, this
enhancement occurs in undriven systems. The intuition is as follows:
In a quantum system with some degree of disorder, localization
suppresses transport at low noise levels. By contrast, at very high
noise levels, decoherence effectively produces a `watchdog effect'
that also suppresses transport. However, at intermediate noise
levels coherence and decoherence can collaborate to produce highly
efficient transport. This enhancement holds even if the noise itself
is purely decohering and can induce no transport on its own. We
investigate a general dephasing model of such environment-assisted
quantum transport (ENAQT) and apply that model to the
Fenna-Matthews-Olson complex and binary trees. For the
Fenna-Matthews-Olson complex we note that the pure-dephasing model
is highly idealized, since it ignores exciton relaxation and
temporal/spatial correlations in the environment. Within these
limitations, we show that the interplay between coherence and
decoherence leads to maximally efficient transport at a noise level
corresponding to ambient temperature.

\emph{Master equations for quantum transport.} The tight-binding
Hamiltonian for an interacting $N$-body system in the presence of a single excitation
is given by \cite{MayBook}:
\begin{equation}
H_{\mathrm{S}}=\sum_{m=1}^{N}\epsilon_{m} \vert m \rangle \langle m \vert
+\sum_{n<m}^{N}V_{mn}(\vert m \rangle \langle n \vert + \vert n \rangle \langle m \vert).
\label{FreeHamiltonian}
\end{equation}%
This Hamiltonian applies to a
large class of quantum transport systems such as excitons and charges in
molecular crystals and quantum dots \cite{Cho05,Nazir05}.
The states $|m\rangle$ denote the excitation being at site $m$.
The site energies and two-body interactions are given by $%
\epsilon _{m}$ and $V_{mn}$ respectively. The site energies, or
static disorder, can be due to different local environments of otherwise
identical molecules or due to fabrication imperfections of engineered
structures. For chromophores, the coupling is mediated by the Coulomb
interaction (F\"{o}rster coupling) or electron exchange (Dexter coupling).
It is appropriate to study the dynamics only in the \emph{single
exciton manifold}, spanned by the states $|m\rangle$. This is
because in the absence of light-matter interaction events the number of
excitons is conserved within the exciton recombination time scale of 1 ns \cite{Owens87},
which is much longer than the usual time scales of the Hamiltonian\ (\ref{FreeHamiltonian}) \cite%
{Ritz01,Adolphs06,Mohseni08}.

A multichromophoric system interacts with the surrounding environment,
such as the solvent or the protein, which is usually a macroscopic system
with many degrees of freedom. This coupling leads to
irreversible dynamics which is characterized by relaxation of
an exciton from a high- to a low-energy state and dephasing of coherences.
At ambient temperature and in common photosynthetic complexes
the energy relaxation of an exciton occurs on a time scale of $\sim$ 1 ps,
while dephasing occurs on a time scale of $\sim$ 100 fs \cite{Leegwater96}.
Thermal fluctuations of the environment couple
to the chromophores by the electron-phonon Hamiltonian:
\begin{equation}
H_{\mathrm{SB}}(t)=\sum_{m}q_{m}(t)\vert m \rangle \langle m \vert,
\label{PhononBathHamiltonian}
\end{equation}%
where the $q_{m}(t)$ describe stochastic bath fluctuations.
Here, we consider only diagonal fluctuations which are typically larger than fluctuations of the
inter-molecular couplings \cite{Adolphs06,Cho05}.
To a certain approximation, the decoherence part of the resulting equation of motion for
a multi-level system in the presence of Markovian fluctuations is dominated by
pure dephasing \cite{Haken73,Leegwater96}. This is
especially the case at high temperatures. The Liouville-von Neumann equation
for the system when averaging over the fluctuations is
$\dot{\rho}(t)=-\frac{i}{\hbar }\langle \lbrack H_{\mathrm{S}}+
H_{\mathrm{SB}}(t),\rho(t)]\rangle$. The random variables
$q_{m}(t)$ are taken to be unbiased Gaussian fluctuations, with
$\langle q_{m}(t)\rangle=0$ and a two-point correlation function
\cite{Cho05,Adolphs06,Leegwater96}:
\begin{equation}
\langle q_{m}(t)q_{n}(0)\rangle =\delta _{mn}\delta (t)\gamma_{\phi},  \label{CorrelationFunction}
\end{equation}%
where $\gamma _{m}$ is a site-dependent rate.
First, we assumed that fluctuations at different sites are uncorrelated.
Second, we assumed that the phonon correlation time
is small compared to the system timescales, an assumption that is justified at room temperatures
where the phonon correlation time is estimated to be below 50 fs \cite{Leegwater96}.
Finally, the correlator is assumed to be site-independent,
so all chromophores experience the same coupling strength to the environment, $\gamma_{\phi }$.
With these assumptions, one obtains the Haken-Strobl equation for the density operator
in the Schr\"{o}dinger picture as \cite{Haken73}:
\begin{equation}
\dot{\rho}(t)=-\frac{i}{\hbar }[H_{\rm S},\rho (t)]+L_{\phi }(\rho (t)),
\label{MasterEq}
\end{equation}
where the pure-dephasing Lindblad operator is given by:
\begin{equation}
L_{\phi }(\rho (t))=\gamma _{\phi }\sum_{m}[A_{m}\rho (t)A_{m}^{\dagger }-%
\frac{1}{2}A_{m}A_{m}^{\dagger }\rho (t)-\frac{1}{2}\rho
(t)A_{m}A_{m}^{\dagger }].  \label{LindbladSuperoperator}
\end{equation}
with the generators $A_{m}=|m\rangle \langle m|$ and a pure dephasing rate
is given by $\gamma _{\phi }$. This Lindblad equation leads to exponential
decay of all coherences in the density operator.

\emph{Energy transfer efficiency and transport time.} There are
several possible ways to measure or quantify the success rate of an
energy transfer
process, such as energy transfer efficiency and transfer time \cite%
{Ritz01,Sener04,Muelken06,Castro07}. First, in order to account for exciton
recombination and exciton trapping, we augment the Hamiltonian (\ref%
{FreeHamiltonian}) with anti-Hermitian parts \cite{MukamelBook,Mohseni08},
\begin{eqnarray}
H_{\mathrm{recomb}}&=&-i\hbar \Gamma \sum_{m} \vert m \rangle \langle m \vert, \\
H_{\mathrm{trap}}&=&-i\hbar\sum_{m}\kappa _{m} \vert m \rangle \langle m \vert.  \label{Non-HermitianHamiltonian}
\end{eqnarray}%
The exciton recombines with a rate $\Gamma $ at every site and is trapped
with a rate $\kappa _{m}$ at certain molecules \cite%
{Leegwater96,Gaab04,Sener04,Muelken07}. The probability that the exciton is
successfully captured at a target site $m$ within the time interval $[t,t+dt]
$ is given by $2 \kappa _{m}\langle m|\rho (t)|m\rangle dt.$ Thus, the
efficiency can be defined as the integrated probability of trapping at
multiple sites as:
\begin{equation}
\eta =2\sum_{m}\kappa _{m}\int_{0}^{\infty }dt~\langle m|\rho (t)|m\rangle .
\label{ETE}
\end{equation}%
In general, the efficiency is reduced by finite exciton lifetimes
($\sim$ 1 ns). Another relevant measure for a quantum transport process is
the average transfer time defined as:
\begin{equation}
\tau =\frac{2}{\eta }\sum_{m}\kappa _{m}\int_{0}^{\infty }dt~t~\langle
m|\rho (t)|m\rangle .  \label{TransferTime}
\end{equation}%
The efficiency of quantum transport elucidates the shorter
time scales given by the trapping/recombination rates. This
approach differs from approaches that consider the limiting
distribution of site populations \cite{Childs02}. For example, the
limiting distribution of a pure-dephasing master equation, such as
Eq.~(\ref{MasterEq}) (i.e.~without the trapping/recombination part),
is an equal population of all sites, which is the same as for a
classical random walk on a regular graph \cite{Childs02}.
The efficiency captures the physically relevant shorter time scales.

\emph{Environment-assisted quantum transport (ENAQT).} The
efficiency of quantum transport in an open system can be
substantially enhanced by the interaction with a fluctuating environment. The master equation (\ref%
{MasterEq}) is a specific example of a large class of transport master
equations representing site-to-site hopping situations. As noted above, the
Hamiltonian part of these master equations has a diagonal part representing
the energies of the individual sites, while the off-diagonal part represents
hopping terms. The open-system Lindblad operators in the master equation are
dominated by terms that dephase the system in the site basis. Relaxation,
another possible non-unitary contribution to the quantum transport involving
energy exchange with the environment, is rather slow compared with dephasing.
The dephasing rate is slow at low temperatures, and fast at high
temperatures. Based on fundamental physical principles, we can make the
following set of phenomenological predictions for such decohered quantum
evolution. As will be seen, these predictions are borne out by the
simulated behavior of the Fenna-Matthews-Olson complex and in binary trees.
We predict that the same generic behavior will hold for decohered quantum
walks in general.

At low temperatures, the dynamics is dominated by coherent hopping. Because
of the variation in the energy levels of different sites and in the strength
of the hopping terms, the system is disordered and exhibits quantum
localization \cite{Anderson58}. The degree of localization depends on the
variation in the energies: for small variation, the system should exhibit
weak localization, and for large variation, strong localization should take
place. Note that the characteristic behavior of quantum localization can
occur even in a system with only a few sites \cite{Henry06}. In this case,
localization can be thought of in terms of energy conservation: an excitonic
state originally localized at an initial site is a superposition of energy
eigenstates that exhibits only a slight overlap with an excitonic state
localized at a final state with significantly different site energy. As a
result, coherent hopping on its own has a low efficiency for transporting an
excitation from one site to another with significantly different site energy.

As the temperature rises, dephasing comes into play. At first, it might seem
that dephasing in the site basis can have no role in enhancing transport,
as this form of noise induces no transport on its own. A moment's
reflection, however, reveals that this expectation is incorrect.
Localization is caused by coherent interference between paths; if that
coherence is destroyed, then the localization effect is mitigated. Coherence
causes an excitation to become `stuck.' It might oscillate back and forth between
a few sites that are strongly coupled and have similar energies, but the
exciton will never venture far afield. By destroying the coherence of the
beating, dephasing also destroys the localization and allows the exciton
to propagate through the system. This phenomenon, by which decoherence
enhances transport, affects all such hopping systems.

When the dephasing rate grows larger than the terms of the system
Hamiltonian we expect transport to be suppressed again. This
suppression of transport by high dephasing can be thought of as an
example of the watchdog (quantum Zeno) effect: rapid dephasing at a
rate $\gamma _{\phi }$ in the site basis is equivalent, so far as
the system is concerned, to being measured repeatedly in the site
basis at time intervals $\approx \gamma _{\phi }^{-1}$. The watchdog
effect will then suppress transport away from the initial site.
While these predictions hold for a large class of transport systems,
we study these effects for three different systems including a
two-chromophore system, the Fenna-Matthews-Olson complex, and a
binary tree structure.

\emph{Quantum transport in a two chromophore system.} A
particularly simple and illustrative example is to study quantum
transport in a system of two sites without trapping and recombination:
a particle hops from site 1 to 2
with a significant energy mismatch between 1 and 2. With $\vert
1\rangle $ and $\vert 2\rangle $, the states where the exciton is
localized at site 1 and 2, respectively, the Hamiltonian for such a
system can be written $H=\epsilon /2(\vert 1\rangle \langle 1\vert
-\vert 2\rangle \langle 2\vert )+V/2(\vert 1\rangle \langle 2\vert
+\vert 2\rangle \langle 1\vert )$, where $\epsilon $ is the energy
mismatch between 1 and 2, and $V$ is the strength of the hopping
term. As usual, we define the Larmor frequency $\hbar \Omega =\sqrt{\epsilon ^{2}+V^{2}}$.
The coherent evolution of the system,
starting from site 1, is simply a rotation about an axis displaced by an
angle $\theta=\sin ^{-1}(V/\hbar \Omega)$ from the $z$-axis in the $x-z$ plane. The maximum
probability of finding the system at site 2 is $\sin ^{2}2\theta $, and the
average probability of finding it there is $\sin ^{2}\theta $. If the energy
mismatch is sufficiently large, substantial hopping does not occur and
the system remains localized at site 1.

In the presence of decoherence, the system obeys the Bloch equation. Pure
dephasing corresponds to a Lindblad operator $\sqrt{\gamma _{\phi }}(\vert
1\rangle \langle 1\vert -\vert 2\rangle \langle 2\vert )$, where $\gamma
_{\phi }=1/T_{\phi }$. The conventional Bloch analysis now holds. The system,
instead of
remaining localized at site 1, gradually diffuses, ultimately becoming a uniform
mixture of $\vert 1\rangle $ and $\vert 2\rangle $.
In the equilibrium state
the system has a $50\%$ chance of being found at site 2. The diffusion
process can be thought of as a random walk on the Bloch sphere with step
length $\theta $ and with time per step $\gamma _{\phi }^{-1}$. Accordingly,
the system must perform $\approx (\pi /\theta )^{2}$ steps and the diffusion
time is $\tau_{\rm diff} \approx (\pi /\theta )^{2}\gamma _{\phi }^{-1}$ to reach a steady
state. For a system with more than two sites, the transport will be more
complicated. Nevertheless, we still expect the transport rate to increase in
direct proportion to the inverse of the individual site decoherence time.
This is indeed true if the decoherence time does not substantially exceed
the time scales defined by the transport terms in the Hamiltonian and the
energy mismatch from site to site. This fact supports the second prediction
of environment-assisted quantum transport.

In the case of rapid dephasing, $\gamma _{\phi }>\Omega $, the angle $\phi $
that the system precesses before being decohered is $\approx \Omega /\gamma
_{\phi }$. The probability of remaining in site $1$ becomes $\cos
^{2}\phi \approx 1-(\Omega /\gamma _{\phi })^{2}$. The system essentially
performs a biased random walk with step size $\phi $ and an average time per
step of $(\gamma _{\phi }/\Omega )^{2}\gamma _{\phi }^{-1}=\gamma _{\phi
}/\Omega ^{2}$. In time $t$, the system diffuses by an angle $\Omega \sqrt{%
t/\gamma _{\phi }}\cdot \Omega /\gamma _{\phi }=\Omega
^{2}t^{1/2}\gamma _{\phi }^{-3/2}.$ In the case that the system has
more than two states, we still expect this analysis to hold, taking
$\gamma _{\phi }$ to be the dephasing rate and $\Omega $ to be an
average eigenfrequency. This supports our third prediction: as the
dephasing rate grows larger than the Hamiltonian energy scale, the
transport rate is suppressed by a polynomial in the dephasing rate.
The system will obviously converge to
the same statistical mixture as mentioned above, albeit on a long
and in some cases physically irrelevant time scale.
These general properties of ENAQT are also observed by the simulations of
the FMO complex and binary trees, where we include recombination and
trapping.

\emph{Quantum transport in Fenna-Matthews-Olson protein complex.}
The Fenna-Matthews-Olson protein of the green
sulphur bacterium \textit{Chlorobium tepidum} \cite{Engel07,Cho05,
Mueh07} is a trimer in which each of the three subunits has seven chlorophyll
molecules spatially arranged within a distance of several nm \cite%
{Li97}. The three subunits can be treated independently from each other.
The FMO complex transfers excitation energy from the chlorosomes,
the main light-harvesting antennae, to a reaction center where a charge
separation event and subsequent biochemical reactions occur. In analogy to
the two-level system discussed above, we expect the same environment-assisted
dynamical behavior in the FMO complex monomers. The dynamics of a single excitation
is governed by a Hamiltonian of the form Eq.~(\ref{FreeHamiltonian}) for seven sites with a
distribution of site energies and inter-site couplings as given in Ref.~\cite%
{Cho05}. The chromophoric F\"{o}rster couplings are up to 100 cm$%
^{-1}.$ The chlorophyll transition frequencies are shifted by the
electrostatic protein environment, resulting in site-dependent
electrochromic shifts of up to 300 cm$^{-1}$ \cite{Adolphs06,Mueh07}.
Fluctuations of the protein in the solvent lead to fluctuations of the
transition frequency of the chlorophyll molecules and therefore to loss of excitonic
phase coherence. We use the master Eq.~(\ref%
{MasterEq}) with a site-independent dephasing rate $\gamma _{\phi }$
according to the Haken-Strobl model. This approach is not the
standard method for describing decoherence effects within the FMO
complex. Usually, relaxation \cite{Cho05,Adolphs06,Mohseni08} and
spatial correlations \cite{Adolphs06} are included. In fact, a more
complete description would also involve a phonon bath with memory
effects. However, the Haken-Strobl model has already been used to
describe the quantum dynamics of certain chromophoric arrays
\cite{Gaab04,Leegwater96} and is expected to deliver insight into
the high temperature behavior of the FMO protein.

\begin{figure}[tbp]
\begin{center}
\includegraphics[scale=1.15]{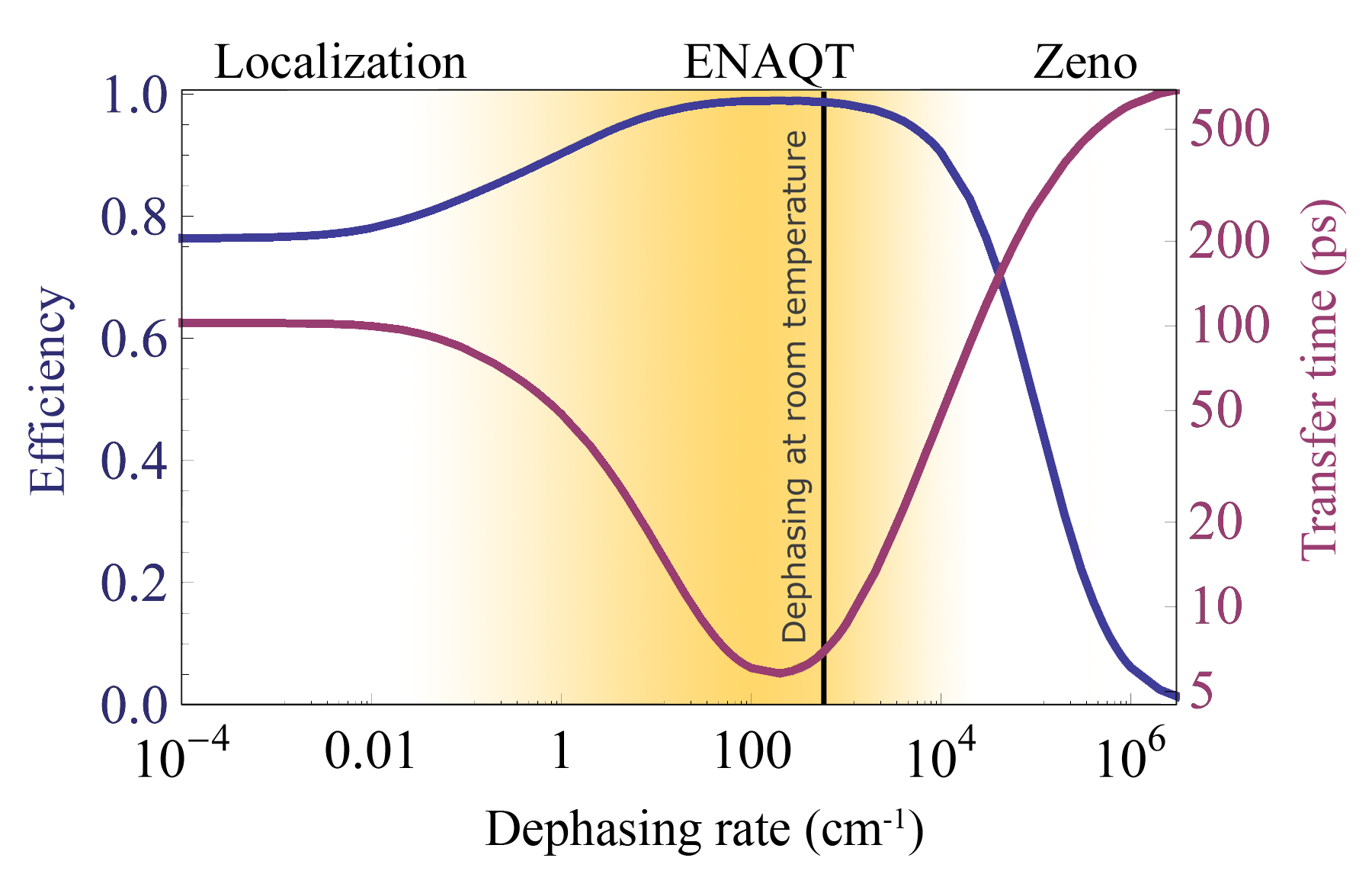} %
\includegraphics[scale=1]{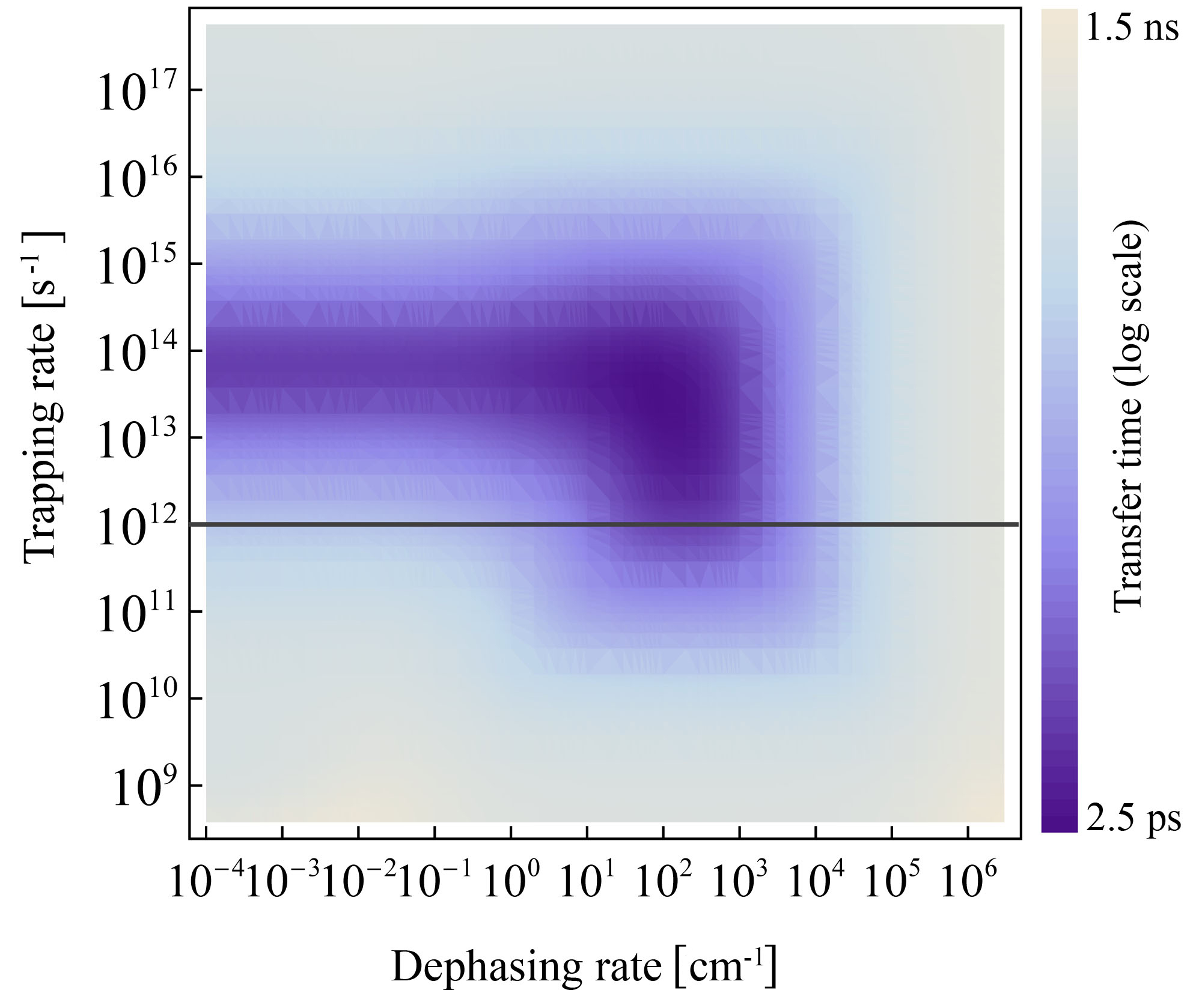}
\end{center}
\caption{(Upper panel) Efficiency (blue) and transfer time (red) as
a function of the pure-dephasing rate is demonstrated for the
Fenna-Matthew-Olsen complex. A clear picture of the three dephasing regimes is obtained:
from left to right, the fully quantum regime which is dominated by intrinsic static disorder in the system Hamiltonian;
the ENAQT regime (qualitatively indicated by the yellow color gradient),
 where unitary evolution and dephasing collaborate with the result of increased efficiency;
finally, the quantum Zeno regime, where strong dephasing suppresses the quantum transport. As a guide to the eye,
the estimated dephasing rate at room temperature for
the FMO complex spectral density (see text) is drawn. The trapping rate is $\protect%
\kappa_{3}=1$ ps$^{-1}$. (Lower panel) Transfer time as a function
of dephasing rate and trapping rate $\protect\kappa _{3}$ is
illustrated on a log scale. The upper panel is indicated as a
horizontal line.} \label{figFMO}
\end{figure}

The initial state for our simulation of the system is a statistical mixture
of localized excitations at sites 1 and 6, the chlorophyll molecules that
are close to the chlorosome antenna. In the FMO complex, chromophore 3 is in
the vicinity of the reaction center \cite{Li97,Adolphs06,Mueh07}. Thus, one can
assume that chlorophyll 3 is the main excitation donor to the reaction
center \cite{Mohseni08}. The precise transfer rate to the reaction center is
not fully characterized. Yet, based on typical transfer rates in
chromophoric complexes with similar inter-molecular distances, we estimate
it to be $\kappa _{3}=1\;\mathrm{ps}^{-1}$ \cite{Mohseni08}. Thus, the
efficiency of energy transfer according to Eq.~(\ref{ETE}) becomes $\eta
=2\kappa _{3}\int_{0}^{\infty }dt\langle 3|\rho (t)|3\rangle .$

In Fig.~\ref{figFMO} (upper panel) the efficiency of transfer and the
transfer time is given as a function of the dephasing rate $\gamma _{\phi }$.
At low dephasing, purely quantum mechanical evolution leads to an
efficiency of around 80\%. With increasing dephasing the efficiency increases
considerably, up to 94\%, where it
approximately remains constant for a range of $\gamma _{\phi }$ of one order
of magnitude. For stronger dephasing the efficiency is slowly suppressed
again, delocalization is destroyed, and the overlap with the target site
vanishes. The transfer time is 75 ps in the fully quantum limit and improves
significantly to 7 ps in the intermediate ENAQT regime. For large dephasing,
the transfer slows down to 500 ps, the same order of magnitude as the
excitation lifetime: the exciton is more likely to recombine than to be trapped.

One can estimate the dephasing rate as a function of
temperature by employing a standard system-reservoir model \cite{BreuerBook}.
In this context, the spectral density is given by
$J(\omega )=\sum_i \omega_i^2 \lambda_i^2 \delta (\omega - \omega_i)$, where
$\omega_i$ are frequencies of the harmonic-oscillator bath modes and $\lambda_i$
are dimensionless couplings to the respective modes.
In the continuum limit, we assume an Ohmic spectral density with cutoff,
$J(\omega )=\frac{E_{\mathrm{R}}}{\hbar \omega _{c}}\omega \exp
(-\omega /\omega _{c})$. For the FMO complex, the reorganization energy is
found to be $E_{\mathrm{R}}=35\mathrm{\ cm}^{-1}$ \cite{Cho05} and the
cutoff $\omega _{c}=150\;\mathrm{cm}^{-1}$, inferred from Fig.~2 in Ref.~\cite{Adolphs06}.
In the Markovian regime, the dephasing rate $\gamma_{\phi}$ is given as the
zero-frequency limit of the Fourier transform of the bath correlator.
As a result, $\gamma_{\phi}$ is found to be proportional to the temperature and the derivative
of the spectral density at vanishing frequency, $\gamma _{\phi }(T)=2\pi \frac{kT}{\hbar}
\left.\frac{\partial J(\omega )}{\partial \omega }\right|_{\omega =0}$ \cite{BreuerBook, Gilmore08}.
For the above spectral density the rate turns out to be
$\gamma _{\phi }(T)=2\pi \frac{kT}{\hbar}\frac{E_{\mathrm{R}}}{\hbar \omega _{c}}$.
This gives a rough estimate for the dephasing rate
at room temperature of around 300 cm$^{-1}$, which is indicated in Fig.~\ref%
{figFMO}. Hence, the natural operating point of the FMO complex is estimated to be
well within the regime of ENAQT, where the dephasing introduced by a fluctuating environment
enhances the energy transfer efficiency. In Fig.~\ref{figFMO} (lower panel),
the transfer time is shown as a function of dephasing rate and trapping time
$\kappa_{3}$. An optimal region with respect to dephasing rate and trapping rate
is obtained.

\begin{figure*}[tbph]
\includegraphics{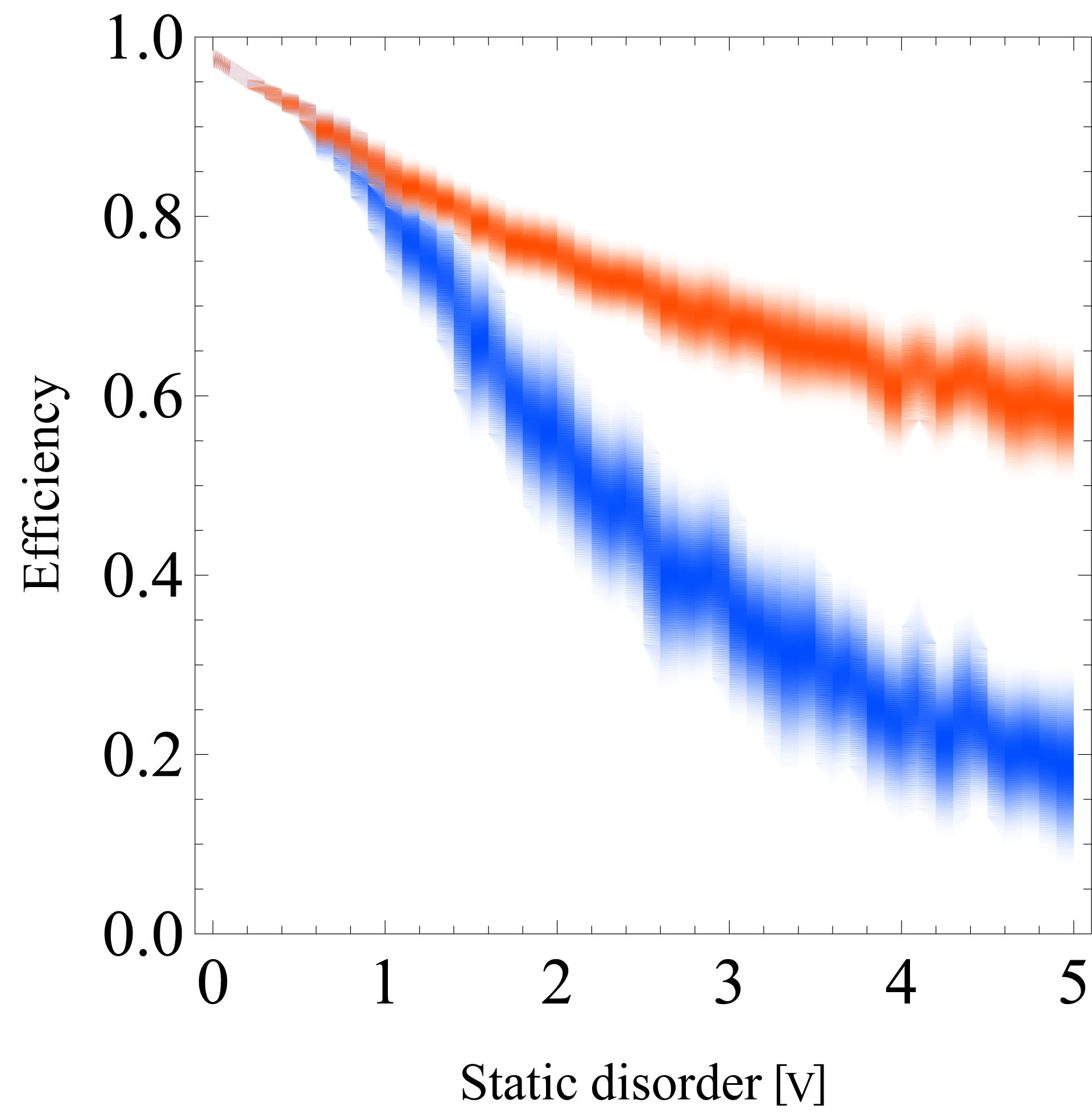} \hspace{1cm}
\includegraphics{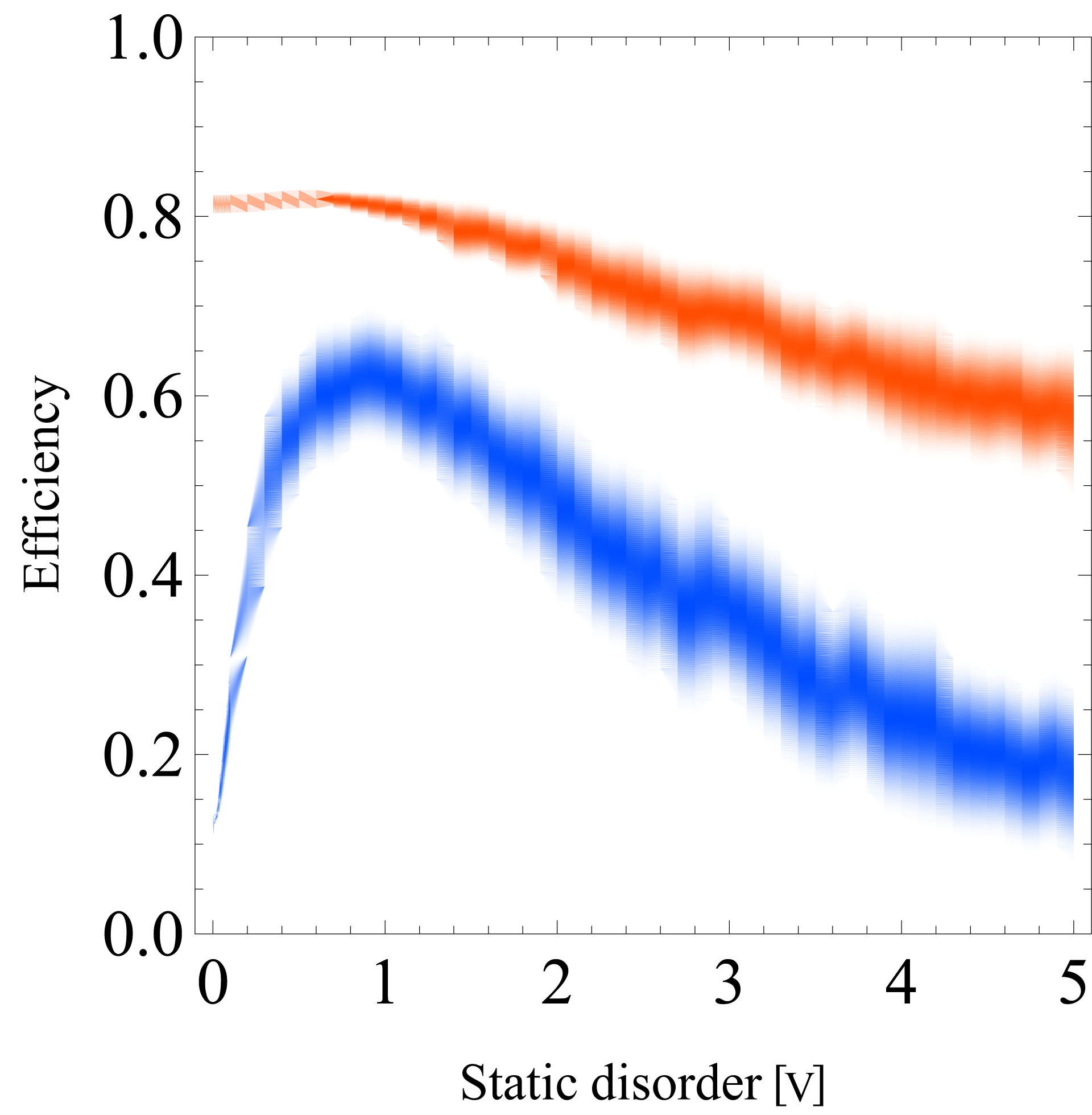}
\caption{The efficiency as a function of the static
disorder parameter $\delta$ (in units of the coupling $V$)
for a four-generation binary tree. A
coherent superposition (left panel) or a statistical mixture (right
panel) of the outermost branch of the tree was chosen as an initial
state (cf. Fig.~1b, grey sites). In the fully quantum case (blue)
large disorder reduces the efficiency due to quantum localization.
100 randomly sampled graphs were used per data point, leading to a
distribution of efficiencies.
For each statically disordered binary tree the optimal dephasing
rate was calculated. The transport efficiency in the presence of dephasing
with this optimal rate is considerably higher (orange).
 Parameters are $\Gamma =0.005V$ and
$\protect\kappa =2V$.} \label{figBinaryTree}
\end{figure*}

\emph{Quantum transport in binary tree structures.} Binary trees
appear in a wide variety of situations, ranging from
computer science \cite{Knuth97} to quantum physics and quantum
information science \cite{Childs02}. Specifically, they arise in
classical and quantum random walks \cite{Childs02} and certain
molecular structures such as dendrimers \cite{DendrimerReview}. The
effect of static disorder of the site energies in quantum walks on
binary trees has been studied in Refs.~\cite{Keating07,Muelken06}
where it was argued that such disorder diminishes the
exponential speed-up in finding particular target sites, a consequence of
quantum localization. The presence of disorder could restrict the
applicability of binary tree structures for devising quantum algorithms or
for transporting quantum particles. In this section, we demonstrate that
ENAQT also occurs in a statically disordered binary tree structure
leading to a substantial improvement in quantum transport.

The Hamiltonian of a disordered graph of generation $g$ is:
\begin{eqnarray}\label{HamiltonianBinaryTree}
H_{\rm S} &=&\sum_{m=1}^{2^{g}-1}\epsilon _{m}|m\rangle \langle m| \\
&&+V\sum_{m=1}^{2^{g-1}-1}(|m\rangle \langle 2m|+|m\rangle \langle
2m+1|+\mathrm{h.c.}).  \nonumber
\end{eqnarray}%
\newline
The site energies $\epsilon_{m}$ are taken to be normally distributed about
a common value $\epsilon_0$, where the standard deviation of the distribution,
$\delta$, is the characteristic parameter of the static disorder. The hopping strength $V$
is uniform over the full graph and connects the sites as depicted
in Fig.~\ref{figTransport}b for a four-generation graph (15 sites).
In the presence of static disorder, the full Hilbert space
has to be taken into account and consequently a reduction to a quantum walk
on the line, as in \cite{Childs02,Keating07}, is not possible.
Additionally, we include exciton recombination at all the sites and
exciton trapping at the center site 1. Both effects are again modeled by the
anti-Hermitian Hamiltonians $H_{\rm recomb}=-i\hbar \Gamma \sum_{m=1}^{2^{g}-1}|m\rangle \langle m|$
and $H_{\rm trap}=-i\hbar \kappa |1\rangle \langle 1|$, with the recombination rate $\Gamma $
and the trapping rate $\kappa$. This defines
the efficiency of transfer, Eq.~(\ref{ETE}).
Dephasing is taken to be uniform over the whole graph according to the master equation \ref{MasterEq}.
We assume an initial state where all the sites in the outermost branch are
populated in a classical mixture or a coherent superposition.

Fig.~\ref{figBinaryTree} shows the dependence of the transport efficiency on the
characteristic disorder parameter $\delta $ for a fourth-generation
binary tree, initially in a coherent superposition (left)
or a statistical mixture (right). The parameters are $\Gamma=0.005V$ and
$\kappa=2V$. For a given $\delta$ the efficiencies are calculated for 100 randomized graphs.
The purely quantum case (blue) is compared to the
case where the introduction of dephasing leads to an optimal enhancement of
the energy transfer efficiency (orange). In the second case the efficiency of
transport for each statically disordered graph was numerically maximized as
a function the pure dephasing rate.
In the figure, the broadening around the average efficiency indicates one standard deviation.
For the initial state being a coherent superposition of all sites
in the outermost branch, one can clearly see that static disorder leads
to a reduced transport efficiency. This is readily explained by localization,
see discussion above and Ref.~\cite{Keating07}.
Dephasing leads to an average improvement of the transport efficiency
for the binary trees considered here. The average improvement is larger the
more static disorder is in the system.

The behavior in the case of small static disorder changes when a classical mixture
of excitations in
the outermost branch is taken as an initial state. Since the initial state
is not a column eigenstate that preserves the symmetry of the graph, the
quantum limit without static disorder shows only small transport
efficiencies of 20\%. Additional static disorder on average increases the
efficiency to a maximum of 60\% when $\delta /V\approx 1.$ Here, static
disorder creates higher localization at the root of the tree at site 1. For
larger static disorder one obtains the same suppression of the efficiency as
for the coherent initial state. In the presence of environmental fluctuations
one obtains an overall average improvement for all static disorder regimes:
for small static disorder the improvement is 60\%, for intermediate it is
20\%, and for larger disorder it is around 40\%.
In summary, ENAQT is shown to consistently improve the transport efficiency
in binary tree structures, overcoming localization induced by static disorder
of the site energies.

\emph{Conclusion.} Environment-assisted quantum transport is
a fundamental effect which occurs in a wide variety
of transport systems. ENAQT is similar in flavor to stochastic resonance \cite%
{Gammaitoni98}: adding noise to a coherent system enhances a
suitable transition rate. ENAQT differs from stochastic resonance in
that the system whose transition rate is enhanced is undriven and
does not need to be in some strong nonlinear regime. The maximum
efficiency of ENAQT occurs when the decoherence rate is comparable
to the energy scales of the coherent system as defined by the energy
mismatch between states and the hopping terms. By changing the
energy mismatch and the hopping terms, the temperature at which the
maximum transport efficiency occurs can be tuned. In the
Fenna-Matthews-Olson protein complex within the pure dephasing model
and with the spectral density as discussed above, this maximum
occurs at approximately room temperature. 
Recently, results along the lines of this work were presented in \cite{Plenio}.
Further studies are in order that utilize more advanced,
non-Markovian decoherence models and detailed quantum chemistry
calculations.

We would like to acknowledge useful discussions with G.R.~Fleming and C.A.~Rodriguez-Rosario.
We thank the Faculty of Arts and Sciences of Harvard University, the Army Research Office
(project W911NF-07-1-0304), and Harvard's Initiative for Quantum Science and
Engineering for funding.

\bigskip

\end{document}